\documentclass[showpacs,notitlepage,groupedaddress,superscriptaddress,twocolumn,numerical,nobalancelastpage,prl,aps]{revtex4-2}

\usepackage[utf8x]{inputenc}

\usepackage{amsfonts,amsmath,amssymb,stmaryrd}

\usepackage{graphicx}
\graphicspath{{./figures/}}

\usepackage[separate-uncertainty = true,multi-part-units=single]{siunitx}
\usepackage[hidelinks]{hyperref}
\usepackage[nameinlink]{cleveref}
\Crefname{equation}{Eq.}{Eqs.}
\Crefname{figure}{Fig.}{Figs.}
\Crefname{tabular}{Tab.}{Tabs.}
\Crefname{table}{Tab.}{Tabs.}

\usepackage{bm} 
\usepackage[dvipsnames]{xcolor}

\newcommand{\mf}{\color{black}}

\newcommand{\iu}{{i\mkern1mu}}
\newcommand{\dder}{\mathrm{d}}
\newcommand{\deriv}[2]{\frac{\mathrm{d}#1}{\mathrm{d}#2}}
\newcommand{\avg}[1]{\left< #1 \right>}

\newcommand{\nex}{N_\mathrm{th}}
\newcommand{\nc}{N_0}
\newcommand{\nexdot}{\dot{N}|_\mathrm{ev}}
\newcommand{\nmoldot}{\dot{N}|_\mathrm{rel}}

\newcommand{\tev}{\dot{T}|_\mathrm{ev}}
\newcommand{\ts}{\dot{T}|_{N_\mathrm{th}}}
\newcommand{\tNo}{\dot{T}|_{N_0}}
\newcommand{\tevap}{\tau_\mathrm{ev}}
\newcommand{\tcoll}{\tau_\mathrm{coll}}
\newcommand{\alphamol}{\alpha}

\newcommand{\cf}{n_\mathrm{c}}
\newcommand{\ttau}{\tilde{\tau}}

\newcommand{\boltzmann}{k_\mathrm{B}}
\newcommand{\ekin}{E_\mathrm{kin}}
\newcommand{\tc}{T_\mathrm{c}}
\newcommand{\vc}{v_\mathrm{c}}
\newcommand{\tauc}{\tau}
\newcommand{\phic}{\varphi_\mathrm{c}}
\newcommand{\omegad}{\omega_\mathrm{d}}
\newcommand{\sigmad}{\sigma_\mathrm{d}}
\newcommand{\crosscorr}{\mathrm{C}_\varphi}
\newcommand{\maxcc}{\max\left(\crosscorr\right)}
\newcommand{\sigmac}{\sigma}
\newcommand{\sigmacl}{\sigma_\mathrm{l}}
\newcommand{\ds}{d_\mathrm{s}}

\newcommand{\vavg}{\overline{V}}
\newcommand{\limol}{$^6$Li$_2$~}
\newcommand{\vs}{v_\mathrm{s}}
\newcommand{\deltats}{\Delta t_\mathrm{s}}

\newcommand{\bk}{{\boldsymbol{k}}}
\newcommand{\br}{{\boldsymbol{r}}}

\DeclareSIUnit\gauss{G}
\DeclareSIUnit\bohrradii{a_0}

\begin{document}

	\title{Ultracold Bose gases in disorder potentials with spatiotemporal dynamics}
		\author{Benjamin Nagler}
	\affiliation{Department of Physics and Research Center OPTIMAS, Technische Universit\"at Kaiserslautern, Germany}
	
\author{Martin Will}
	\affiliation{Department of Physics and Research Center OPTIMAS, Technische Universit\"at Kaiserslautern, Germany}	
	
	\author{Silvia Hiebel}
	\affiliation{Department of Physics and Research Center OPTIMAS, Technische Universit\"at Kaiserslautern, Germany}
	
	\author{Sian Barbosa}
	\affiliation{Department of Physics and Research Center OPTIMAS, Technische Universit\"at Kaiserslautern, Germany}

	\author{Jennifer Koch}
	\affiliation{Department of Physics and Research Center OPTIMAS, Technische Universit\"at Kaiserslautern, Germany}
	
	\author{Michael Fleischhauer}
	\affiliation{Department of Physics and Research Center OPTIMAS, Technische Universit\"at Kaiserslautern, Germany}	
	
	\author{Artur Widera}
	\email{email: widera@physik.uni-kl.de}
	\affiliation{Department of Physics and Research Center OPTIMAS, Technische Universit\"at Kaiserslautern, Germany}

	\date{\today}

	\begin{abstract}
		We study experimentally the dissipative dynamics of ultracold bosonic gases in a dynamic disorder potential with tunable correlation time. First, we measure the heating rate of thermal clouds exposed to the dynamic potential and present a model of the heating process, revealing the microscopic origin of dissipation {\mf from a thermal, trapped cloud of bosons}. Second, for Bose-Einstein condensates, we measure the particle loss rate induced by the dynamic environment. Depending on the correlation time, the losses are either dominated by heating of residual thermal particles or the creation of excitations in the superfluid, a notion we substantiate with a rate model. Our results {\mf illuminate the interplay between superfluidity and time-dependent disorder and on more general grounds} establish ultracold atoms as a platform for studying spatiotemporal noise and time-dependent disorder.
	\end{abstract}
	
	\maketitle

		Disorder is ubiquitous, and its impact on physical systems has been studied intensely in recent decades~\cite{vojtaDisorder2019,abrahams502010}.
		Most investigations were focused on static disorder, in which single-particle wave transport can be suppressed due to Anderson localization~\cite{andersonAbsence1958,wiersmaLocalization1997a,roatiAnderson2008,billyDirect2008,kondovDisorderInduced2015a,schreiberObservation2015a}, and thermalization is absent in certain interacting systems ~\cite{kondovDisorderInduced2015a,schreiberObservation2015a,imbrieManyBody2016,smithManybody2016a,weiExploring2018}. 
		{\mf Since phenomena like Anderson localization are based on interference, modulating disorder in time has dramatic effects.} Recent studies of dynamic disorder in classical and quantum systems 
		{\mf focussing} on transport, {\mf showed} 
		in stark contrast to the static case, {\mf that it can be} supported~\cite{hanggiArtificial2009,gopalakrishnanNoiseInduced2017} and even accelerated beyond the ballistic regime~\cite{jayannavarNondiffusive1982,leviHypertransport2012}. 
		However, {\mf the interplay between} 
		superfluidity and long-range coherence 
		{\mf with} time-dependent disorder, and dissipation induced by the dynamic environment, have not yet been investigated in experiments. The impact of dynamic disorder is of broad interest, for example, in the context of energy transfer in biological systems~\cite{rebentrostEnvironmentassisted2009,chinNoiseassisted2010}, the electrical conductivity of ionic polymers~\cite{ratnerConductivity1989} and microemulsions~\cite{grestDynamic1986}, chemical reactions~\cite{sendina-nadalBrownian2000}, wave propagation in the sea~\cite{virovlyanskyRay2012}, superconductors~\cite{aronovEffect1994}, and quantum walks~\cite{yinQuantum2008}.
		 Theoretical works on spatiotemporal noise predict a nonequilibrium phase transition~\cite{vandenbroeckNoiseInduced1994,garcia-ojalvoColored1994} which is induced by the random environment.
		For quantum systems, it seems natural to pose the question if there is \textcolor{black}{an extention of preparing nonequilibrium states by spatiotemporal periodic drive~\cite{eckardtColloquium2017,singhQuantifying2019} to the case of {\mf general} broad-band spatiotemporal noise. This novel regime is particularly complicated by the nonlinearity of interacting quantum systems as Bose-Einstein condensates (BECs), giving rise {\mf to collective phenomena such as superfluid flow.}}
		\textcolor{black}{One potential challenge is the unfavorable heating of atomic systems due to energy absorption from the dynamic environment~\cite{eckardtColloquium2017}. }
		The role of dissipation is of general interest in the paradigm of open quantum systems~\cite{breuer2007theoryopen}, which is realized by, e.g., quantum gases coupled to environments with spatiotemporal noise.
		
		Here, we study the nonequilibrium dynamics of ultracold molecular Li$_2$ gases in {\mf tunable} dynamical disorder. We employ a novel scheme to realize a time-dependent optical speckle potential with 
		{\mf variable} correlation time, inspired by a method for the decorrelation of light fields~\cite{Arecchi65}.
		\textcolor{black}{For ultracold, thermal ensembles, we observe the microscopic onset of dissipation for decreasing correlation time, which is well described by a random-walk model in momentum space. For BECs, the disorder additionally creates direct excitations in the superfluid, depleting the superfluid fraction. We model the dissipative dynamics of the quantum gas by an open-system rate model, treating the superfluid excitations in two complementary ways. Importantly, comparison with experimental data suggests a window of correlation times having negligible superfluid excitations, well suited for studies of nonequilibrium dynamics of quantum fluids.}
		
		\begin{figure*}
			\includegraphics{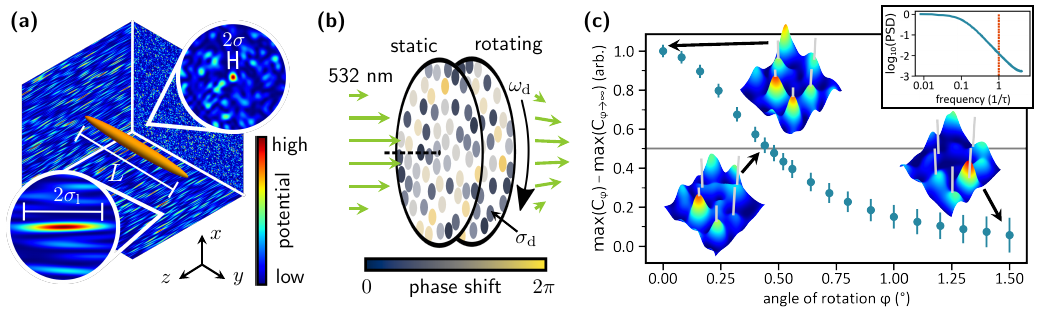}
			\caption{
				(a)~Sketch of experimental realization. Cigar-shaped clouds of \limol molecules with typical size $L\sim\SI{300}{\micro\meter}$ are exposed to an anisotropic speckle potential. 
				(b)~Creation of dynamic speckle. The dots with  size $\sigmad$ represent the random surface of the diffusers, and their colors indicate the magnitude of the phase shift they imprint on incident light. The transmitted light is focused on the cloud.
				(c)~Evolution of a dynamic speckle pattern. Maximum value of the cross-correlation function $\crosscorr$ of the speckle intensity. Error bars mark the uncertainty of a fit that is used to extract the maximum value from $\crosscorr$. Insets show a section of a simulated speckle pattern with $\maxcc$, as indicated by the arrows. Gray lines mark the positions of five distinct peaks in the initial speckle and simplify tracking the evolution of the intensity distribution. The inset plot shows the calculated temporal power spectral density ($\mathrm{PSD}$) of a dynamic speckle (blue, solid line), where the inverse correlation time roughly coincides with the frequency at which $\mathrm{PSD}$ has dropped to $1/100$ of its maximum value at zero frequency. For comparison, we also show the $\mathrm{PSD}$ of a speckle whose mean potential is periodically modulated with frequency $1/\tauc$ (red, dashed).}
			\label{fig:figure1}
		\end{figure*}

		Experimentally, we prepare dilute gases of bosonic \limol Feshbach molecules in a cigar-shaped hybrid magnetic-optical trap (\Cref{fig:figure1}~(a)), for details see Refs.~\cite{Ganger2018,Nagler2020}. 
		The magnetic field \textcolor{black}{close to a magnetic Feshbach resonance at \SI{832.2}{\gauss}~\cite{Zuern2013}} sets the $s$-wave scattering length  $a$ between the molecules and thus their binding energy. Typical thermal (degenerate) samples contain ${>10^5}$ molecules at a temperature of ${T=\SI{590}{\nano\kelvin}}$ (\SI{50}{\nano\kelvin}). 
		A repulsive optical speckle potential~\cite{Goodman2007} at a wavelength of \SI{532}{\nano\meter} introduces the disorder.
		 The typical size of the anisotropic speckle grains is ${\sigma^2 \times \sigma_\mathrm{l}}$ with ${\sigma=\SI{750}{\nano\meter}}$ and ${\sigma_\mathrm{l} = \SI{10.2}{\micro\meter}}$ the correlation lengths along the $x/y$- and $z$-direction. We characterize the strength of the disorder by the spatial average $\vavg$ of the speckle potential at the cloud position.
		
		\textcolor{black}{We create the rotated speckle pattern by transmitting a laser beam through two glass plates with random surface structures, i.e. diffusers, rotated against each other, and focusing the light field onto the atoms (\Cref{fig:figure1}~(b)). Upon rotation, the local phase imprints change significantly, causing the height and position of the interference pattern's speckle grains to change.}
		We quantify the resemblance to the initial speckle intensity distribution $I_{\varphi=\ang{0}}$ by the maximum value of the cross-correlation function~\cite{papoulis1962fourier}, $\maxcc$, with
			\begin{equation}
				\crosscorr(x,y)=\int \dder x'\dder y' I_{\varphi=\ang{0}}(x',y')I_{\varphi}(x'+x,y'+y).
			\end{equation}
		$I_{\varphi}(x,y)$ are two-dimensional intensity distributions in the focal plane for rotation angle $\varphi$ of the diffuser plate, independently measured in a test setup. 
		We define the correlation angle $\phic$ at which $\maxcc$ has dropped to half its initial value, \Cref{fig:figure1}~(c). 
		For rotation at constant angular velocity $\omegad$, the correlation angle translates into a correlation time $\tauc = \phic/\omegad$.
		In the experimental setup, ${\omegad \leq \SI{2100}{\degree\per\second}}$ and ${\phic = \ang{0.6}}$, hence ${\tauc > \SI{285}{\micro\second}}$. Importantly, \textcolor{black}{in contrast} to a periodically driven potential, the temporal power-spectral density of \textcolor{black}{this} dynamic speckle comprises a broad distribution of frequencies, where low-frequency contributions dominate, and the inverse correlation time can be interpreted as a \textcolor{black}{bandwidth or} cut-off frequency (see inset of \Cref{fig:figure1}~(c)). 
		
\begin{figure}
			\includegraphics{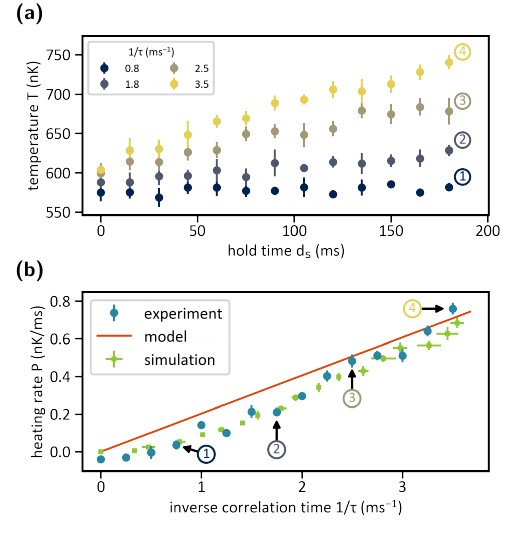}
			\caption{Heating of a thermal ensemble with initial temperature ${T=\SI{590}{\nano\kelvin}}$ in dynamic speckle disorder with ${\vavg=\SI{30.5}{\nano\kelvin}\times \boltzmann}$. (a)~Cloud temperature $T$ versus hold time $\ds$ for various values of $1/\tauc$. (b)~Heating rate $P$ versus inverse correlation time $1/\tauc$. Squares result from the numerical simulation, solid line from the microscopic model. Error bars of experimental data in (b) denotes uncertainty of the fit, other errors the standard deviation of 5~repetitions.}
			\label{fig:figure2}
		\end{figure}

		To study the response of thermal clouds to the dynamic disorder, we prepare samples with \num{3.4e5} molecules with ${a=\SI{1524}{\bohrradii}}$ (\si{\bohrradii} is the Bohr radius) in a trap with harmonic frequencies ${\omega_x,\omega_y,\omega_z=2\pi\times(\num{498}, \num{22.1}, \num{340})~\si{\hertz}}$ at a temperature of ${T = \SI{590}{\nano\kelvin}}$. Following the end of the evaporation ramp, the cloud is allowed to relax for \SI{500}{\milli\second} to ensure thermal equilibrium. In order to minimize excitations in the gas, we increase the potential of the dynamic speckle during a \SI{50}{\milli\second} linear ramp to its final value of ${\vavg/\boltzmann=\SI{30.5}{\nano\kelvin}\ll T}$, where $\boltzmann$ is the Boltzmann constant.
		After a variable hold time ${\ds\leq \SI{180}{\milli\second}}$, the speckle is extinguished during \SI{50}{\milli\second}, and we take an absorption image of the trapped cloud.
		We extract the temperature by fitting a Bose-enhanced Gaussian function~\cite{Ketterle1999} to the integrated column-density distribution. We observe that the cloud temperature $T$ is proportional to the hold time $\ds$ and the slope, i.e., the heating rate ${P=\dder T/\dder \ds}$, grows with increasing $1/\tauc$, see \Cref{fig:figure2}. The heating rate is extracted by fitting a linear function to the data. We compare these results to a numerical simulation of classical, noninteracting point particles with thermal velocity distribution in a dynamic, homogeneous speckle in two dimensions~\cite{supps}. The dimensional reduction is facilitated by the anisotropic speckle, which allows to neglect the much weaker potential gradients along the $z$-axis as compared to the $xy$-plane. The heating rates from this simulation (\Cref{fig:figure2}~(b)) yield good agreement with the experimental data. 
		\textcolor{black}{We conclude} that the heating is intrinsically a single-particle effect, not modified by the elastic molecule-molecule scattering at a rate of \SI{11}{\per\milli\second} or inelastic collisions.	
		\textcolor{black}{Moreover, we develop a microscopic heating model based on a random walk in momentum space for the limiting case ${\boltzmann T \gg \vavg}$, which is realized in the experiment. Single particles travel on almost straight trajectories, and experience "kicks" with momentum change ${\Delta p \ll p}$ from the time-dependent potential.}
		The resulting heating rate is given by
			\begin{equation}
				P = \frac{\vavg^2}{2\boltzmann^2 T \tauc}\gamma,
				\label{eq:heatingrate}
			\end{equation}
		where the constant $\gamma$ corrects for the dimensionality and the trapping potential in each data set independently~\cite{supps}. The model matches the measured heating rates for sufficiently large inverse correlation times (\Cref{fig:figure2}~(b)). 
		Theoretical works on the transport of classical particles in dynamic disorder predict a universal time dependence of the average kinetic energy of a particle ${\ekin\propto t^{2/5}}$~\cite{golubovicClassical1991,rosenbluthComment1992,krivolapovUniversality2012}. The fact that we observe constant heating rates, i.e., ${\ekin\propto t}$, is most likely due to the short observation times, which do not allow to discriminate between linear and sub-linear behavior. Power-law behavior is indeed reproduced in the numerical simulation for sufficiently long observation times.
		
		\begin{figure}
			\includegraphics{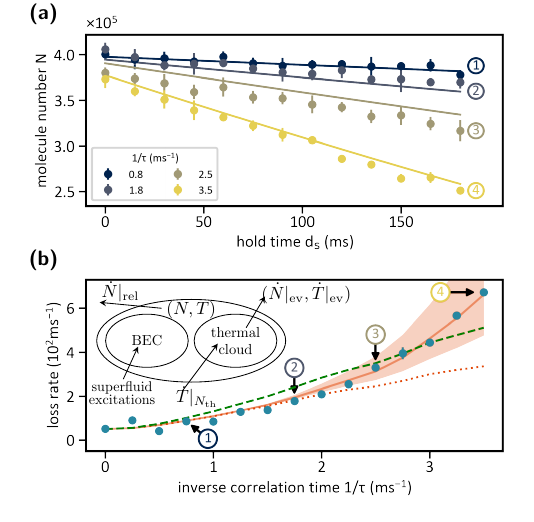}
			\caption{
				Dissipation of a BEC in dynamic disorder. (a)~Total molecule number $N$ versus hold time $\ds$ for various values of $1/\tauc$. Error bars denote the standard deviation of 5~repetitions. Solid lines are from the rate model. (b)~Loss rates versus inverse correlation time $1/\tauc$. Error bars of experimental data points (blue) show the error estimation of the linear fit and are smaller than the marker size for most data points.
				{\mf Lines indicate results from the rate model, including thermal heating and superfluid excitations (solid), thermal heating and particle loss from the condensate (dashed green), or only heating of the thermal cloud (dotted).}
				The shaded area represents a $\pm \SI{20}{\percent}$ variation of $\vs$. The inset illustrates the processes included in the open-system rate model.}
			\label{fig:figure3}
		\end{figure}
	
		\begin{figure*}
			\includegraphics{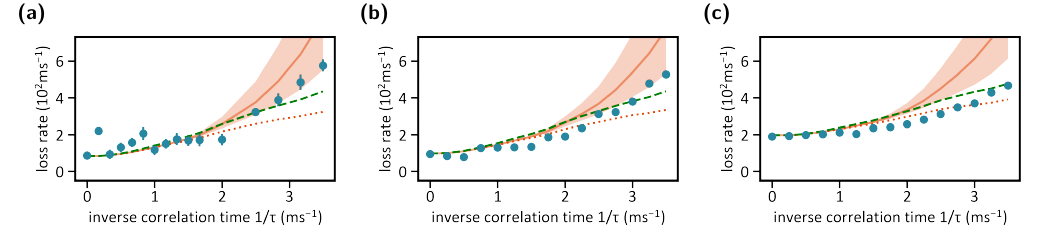}
			\caption{Loss rates of BECs in dynamic speckle for various values of the $s$-wave scattering length $a$. {\mf The allocation of colors and line styles is the same as in \Cref{fig:figure3}~(b).} (a)~${a=\SI{1524}{\bohrradii}}$, ${\mu=\SI{187}{\nano\kelvin}\times\boltzmann=\SI{3.9}{\kilo\hertz}\times h}$ (b)~${a=\SI{1310}{\bohrradii}}$, ${\mu=\SI{173}{\nano\kelvin}\times\boltzmann=\SI{3.6}{\kilo\hertz}\times h}$ (c)~${a=\SI{982}{\bohrradii}}$, ${\mu=\SI{144}{\nano\kelvin}\times\boltzmann=\SI{3.0}{\kilo\hertz}\times h}$. \textcolor{black}{The range of interaction strengths explored is limited for lower interactions by the decreasing collisional lifetime of molecules; and for higher interactions by the emergence of free atoms as the binding energy decreases.}}
			\label{fig:figure4}
		\end{figure*}
		
		In order to study quantum gases in dynamical disorder, we cool samples with ${N=\num{4e5}}$ molecules and scattering length ${a=\SI{2706}{\bohrradii}}$ to ${T=\SI{50}{\nano\kelvin}}$, far below the noninteracting critical temperature
		{\mf of condensation} ${\tc=\SI{245}{\nano\kelvin}}$. Hence, we expect a condensate fraction $>\num{.8}$ and a BEC with chemical potential ${\mu=\SI{250}{\nano\kelvin}\times\boltzmann=\SI{5.2}{\kilo\hertz}\times h}$, where $h$ is Planck's constant. The corresponding time scale ${h/\mu=\SI{190}{\micro\second}}$ is smaller than the experimentally accessible correlation times, and the healing length at the trap center ${\xi=\SI{380}{\nano\meter}}$~\cite{pethick_smith_2008} falls below the correlation lengths. Thus, for these maximum values, the condensate can temporally react to and spatially resolve all changes and details of the speckle potential. The experimental sequence for the exposure to the dynamic speckle is the same as for thermal clouds. Instead of the temperature, we monitor the total molecule number $N$ of the sample, because the large condensed fraction does not allow to extract a temperature from absorption images reliably. 
		We find that the molecule number decreases linearly with $\ds$, and the loss rate $-\dder N/\dder \ds$ grows with $1/\tauc$~(see \Cref{fig:figure3}). 
		We distinguish two main processes contributing to the loss of molecules from the trap. On the one hand, as described before, the dynamic speckle heats the residual thermal component of the gas. 
		The rising temperature causes molecules to transfer from the BEC to the thermal fraction, from which molecules with sufficient energy can evaporate, which in turn cools the sample. On the other hand, the motion of the dynamic speckle creates excitations in the BEC, which again diminishes the condensate fraction because of Landau damping~\cite{pitaevskiiLandau1997}. 
		\textcolor{black}{We model the underlying dynamics by two approaches. The first takes into account the trap but treats superfluid damping in a phenomenological way, whereas the second provides analytic expressions for the particle loss from the condensate fraction in a homogeneous superfluid.}
		
		{\mf Due to the BEC being superfluid, excitations are mainly expected if the typical velocity $\vs$ of the speckle exceeds the local Landau critical velocity ${\vc(\bm{r})=\sqrt{gn_0(\bm{r})/m}}$ in the condensate, where $n_0$ is the condensate density distribution and $g$ the coupling constant~\cite{pethick_smith_2008}. These local quantities are well-defined because, for our parameters, the local-density approximation is valid~\cite{kaganBoseEinstein1996a}.}	We can estimate the largest velocity scale of the speckle from the correlation lengths and time to be ${\vs=\sqrt[3]{\sigmac^2\sigmacl}/\tauc<\SI{6.3}{\milli\meter\per\second}}$, which is below the maximum critical velocity ${\vc(\bm{0})=\SI{13.2}{\milli\meter\per\second}}$ at the center of the condensate. However, \textcolor{black}{because of the Thomas-Fermi density profile~\cite{stringaribec}}, there are always regions with ${\vc(\bm{r})<\vs}$ where excitations are possible. Additionally, inelastic collisions between molecules cause losses, even in the absence of any speckle potential~\cite{petrovWeakly2004}. We capture this interplay between heating, evaporation, and cooling by a set of rate equations
			\begin{align}
				\dot{N} &= \nexdot + \nmoldot \label{eq:ratemodel1} \\
				\dot{T} &= \ts + \tev + (\tNo) \label{eq:ratemodel2}
			\end{align}
		modeling the open quantum system (for details see Ref.~\cite{supps}), which include the processes evaporation from the thermal component ${(\nexdot,\tev)}$, molecular relaxation $\nmoldot$, and heating of the thermal component by the dynamic speckle $\ts$~(see inset of \Cref{fig:figure3}~(b)). 
		We calculate the number of superfluid molecules ${\nc=N\times\cf(T/\tc,N,a)}$ using an expression for the condensate fraction $\cf$, which incorporates the intermolecular interaction and finite size of the system~\cite{xiongCritical2001,supps}.
		\textcolor{black}{We neglect effects of the relatively strong quantum depletion~\cite{xiongCritical2001}, because the depleted density remains superfluid~\cite{millerElementary1962}.}
		The number of thermal molecules is given by ${\nex=N-\nc}$ and we assume the system to be in thermal equilibrium at all times. 
		{\mf In order to include the effect of the speckle potential onto the superfluid molecules, in a first approach, we calculate the fraction $f$ of the ones located in regions of the condensate where ${\vc(\bm{r})<\vs}$. We assume that in addition to thermally excited molecules $\nex$, the condensed particles in the former mentioned area $f \times N_0$ are removed from the system by evaporation. Numerically, we find that $f$ is close to zero below ${\vs/\vc(\bm{0})=\num{.3}}$~\cite{supps}, which roughly coincides with ${1/\tauc\approx \SI{2}{\per\milli\second}}$. }
		
		{\mf This approach obviously neglects the intricate dynamics, interactions, and spectrum of superfluid excitations excitations~\cite{quantumliquids}.}
 		Therefore, in a second approach, we compute the rate of particles transferred from the condensate to the thermal fraction \textcolor{black}{using number-conserving Bogoliubov theory in a speckle with Gaussian-shaped spatiotemporal spectrum \cite{Castin2001,supps}}, contributing another heating term $\tNo$ in \Cref{eq:ratemodel2}.
		For a homogeneous condensate, we find~\cite{supps}
		\begin{align}
			\tNo =& \frac{\eta^2 \bar{V}^2\;  T_c^3 \pi \sigma^3 }{6 T^2 \hbar^2 v_c  \xi\;\sqrt{\sigma^2+v_c^2 \ttau^2  }}  e^u \bigg\{ 2u \Big[ I_{5/4}(u) \label{eq:heatingN0} \\
			&  -I_{3/4}(u)+I_{1/4}(u) -I_{-1/4}(u) \Big] +I_{1/4}(u) \bigg\}, \notag
		\end{align}
		where $I_{\nu}(u)$ are modified Bessel function of the first kind \cite{abramowitz2008}, $u$ is defined as $u = {(\sigma^2 + v_c^2 \ttau^2 )^2}/{16 \xi^2 v_c^2 \ttau^2}$ and $\ttau = \tau/\sqrt{\log{2}}$. {\mf In oder to adopt the homogeneous theory to the inhomogeneous experimental system, we use the fit parameter $\eta = 1/20$ but evaluate \Cref{eq:heatingN0} with the mean superfluid density.}
		We solve \Cref{eq:ratemodel1,eq:ratemodel2} numerically to obtain the time dependence of the particle number and compare the results to the experimental data in \Cref{fig:figure3}. 
	    \textcolor{black}{Both models} reproduce the measured loss rates closely. 
	    \textcolor{black}{Molecular relaxation is included via the relaxation rate} $\alphamol$ such that the loss rate in the static speckle matches the measured one; we find agreement with previously reported values~\cite{supps,cubizollesProduction2003a,jochimPure2003}. 
	    For relatively long correlation times ${1/\tauc \lesssim \SI{2}{\per\milli\second}}$, the losses due to superfluid excitations are negligible, and the loss rates are well captured merely by the heating of the thermal cloud~(dotted line in \Cref{fig:figure3}~(b)). 
		In the case ${1/\tauc \gtrsim \SI{2}{\per\milli\second}}$, both loss mechanisms contribute significantly.
		\textcolor{black}{Reducing the interaction strength, the phenomenological rate model systematically overestimates the loss rate 
		{\mf by assuming immediate depletion of condensate atoms in the region $v_c({\vec r})<v_s$}
		(see \Cref{fig:figure4} and Ref.~\cite{supps}), while the model computing the excitation rate from the condensate yields good agreement for all interaction strengths {\mf with one common fit parameter}.}
	
		Our studies indicate a regime, where quantum fluids are shielded from direct superfluid excitations even for a broad-band excitation, prevailing for a broad range of interaction strengths.
		The tight control over correlation times points toward future studies of transport in time-dependent disorder both for classical and quantum systems with strong interactions. 
		
	\section*{Acknowledgements}
		We thank Hans Kroha, and Axel Pelster for fruitful discussions and Maximilian Kaiser for carefully reading the manuscript. This work was supported by the Deutsche Forschungsgemeinschaft (DFG, German Research Foundation) via the Collaborative Research Center SFB/TR185 (Project No. 277625399). J.K. and M.W. were supported by the Max Planck Graduate Center with the Johannes Gutenberg-Universität Mainz (MPGC).
	
	
	\bibliography{bibliography,bibliographyauto}{}

	\appendix*
	\clearpage

\section{Supplementary Material}
In the following, details on the experimental procedure, the theoretical models and additional data are given. 

	\subsection{Experimental procedure}
	
		\begin{figure}[b]
			\includegraphics{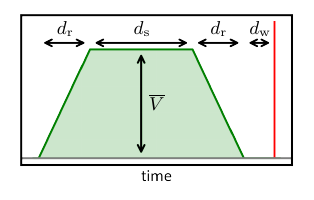}
			\caption{Experimental sequence. Following the end of the evaporation ramp and a hold time of \SI{500}{\milli\second}, the dynamic speckle is ramped up linearly during ${d_\mathrm{r}=\SI{50}{\milli\second}}$. After a variable hold time ${\ds<\SI{180}{\milli\second}}$, the speckle is slowly extinguished and we take an in-situ absorption image (red line) of the cloud after a waiting time ${d_\mathrm{w}=\SI{30}{\milli\second}}$.}
			\label{fig:supp_figure1}
		\end{figure}
		
		We prepare quantum gases in the BEC-BCS crossover regime by forced evaporative cooling of fermionic $^6$Li atoms in an equal mixture of the two lowest-lying Zeeman substates of the electronic ground state $^2\mathrm{S}_{1/2}$. Evaporation takes place in a hybrid magnetic-optical trap at a magnetic field of $\SI{763.6}{\gauss}$ on the repulsive side of a Feshbach resonance centered at \SI{832.2}{\gauss} \cite{Zuern2013}, where atoms of opposite spin form bosonic molecules that eventually condense into a BEC. After evaporation, the sample is held at constant trap depth for \SI{300}{\milli\second} to ensure thermal equilibrium before the magnetic field is linearly ramped to its final value during \SI{200}{\milli\second}. At this point, the dynamic speckle is introduced by ramping the laser power linearly during \SI{50}{\milli\second} to its final value (\Cref{fig:supp_figure1}). The laser power is held constant for a variable time $\ds$ and subsequently extinguished. After a waiting time of \SI{30}{\milli\second}, we employ resonant high-intensity absorption imaging \cite{Reinaudi2007} to extract the column density distribution in the $y$-$z$-plane.
		For thermal clouds, the temperature is determined by fitting a Bose-enhanced Gaussian function to the density distribution. In the case of BECs, we estimate the sample temperature to be $T=\SI{50(25)}{\nano\kelvin}$ by ramping the magnetic field to \SI{680}{\gauss} prior to imaging and fitting a bimodal density distribution \cite{naraschewskiAnalytical1998a}.
		
		The hybrid trap consists of an optical dipole trap and a magnetic saddle potential, which provides weak (anti-) confinement in ($z$-) $x$- and $y$-direction, whereas the optical trap strongly constrains the cloud along $x$ and $z$. Since the saddle potential is an accessory to the magnetic field used to address the Feshbach resonance, its curvature depends on the field magnitude. The trapping frequencies and other relevant parameters for all presented experimental data are listed in \Cref{tab:parameters}.
		
		\begin{table}[b]
			\begin{tabular}{l | r | r | r | r | r}
				magnetic field (G) & 700.0 & 720.0 & 730.0 & 763.6 \\
				\hline
				$\omega_y/2\pi$ (\si{\hertz}) & 21.7 & 22.0 & 22.1 & 22.6 \\
				$a$ ($a_0$) & 982 & 1310 & 1524 & 2706 \\
				$N(0)$ ($10^3$) & 288 & 325 & 345 & 406 \\
				$\alphamol$ $(10^{-13}\si{\centi\meter\cubed\per\second})$ & 2.9 & 1.0 & 1.2 & 0.65
			\end{tabular}
			\caption{Overview of parameters for different magnetic fields. Scattering lengths taken from \cite{Zuern2013}. $N(0)$ is the initial molecule number used for the solution of \Cref{eq:ratemodelsupps1,eq:ratemodelsupps2}. $\alphamol$ is the molecular relaxation rate.}
			\label{tab:parameters}
		\end{table}
		
		The speckle potential is created by passing a laser beam of wavelength \SI{532}{\nano\meter} through two diffusive plates (Edmund Optics 47-988 and 47-991) and focusing the light, using an objective with numerical aperture \num{0.29}, onto the atoms. They experience a repulsive and spatially random dipole potential $V$, which we characterize by its spatial average $\vavg$ at the focal point of the objective. The typical grain size of the speckle is given by the Gaussian-shaped autocorrelation function of the potential with $1/e$ widths (correlation lengths) ${\sigmac=\SI{750}{\nano\meter}}$ transversely to and ${\sigmacl=\SI{10.2}{\micro\meter}}$ along the beam propagation direction. As the speckle beam has a Gaussian envelope with waist \SI{440}{\micro\meter}, the average potential is inhomogeneous across the spatial extension of the cloud. We use a motorized rotation stage (OWIS DRTM 65-D35-HiDS) to rotate one of the circular diffusers around its principal axis.
		{As a consequence, the rotation speed and hence the phase shift imprinted onto the light field depends on the distance from the rotation axis. This renders the correlation time $k$-vector dependent. However, the light-field distribution is imaged onto the plane of the atoms, which is deep in the Fraunhofer limit. Thus at every position of the atoms, all $k$-vectors contribute to the interference, yielding a Gaussian correlation in space and time with the correlation length and time as given in the manuscript.}
		
	\subsection{Dynamical Speckle Potential}
		The static speckle potential is created by transmitting a laser beam through a glass plate with a random surface structure, i.e., a diffuser. The diffuser imprints a phase pattern whose spatial variation is characterized by the correlation length ${\sigmad \approx \SI{20}{\micro\meter}}$ of the surface structure (\Cref{fig:figure1}~(b)). 
		By focusing the beam, all partial waves with random phases interfere and create a static speckle pattern with correlation length $\sigmac$ in the focal plane. 
		The speckle is rendered dynamic by adding a second, similar diffuser directly after the first one, which is mounted in a motorized rotation stage. 
		Upon rotation of the second diffuser, the details of the imprinted phase pattern are altered significantly once the local displacement of the diffuser is comparable to $\sigmad$. 
		As a consequence, the height and position of the speckle grains change until the intensity distribution bears no resemblance to its initial state before rotation, see~\Cref{fig:figure1}~(c).

	\subsection{Numerical simulation of classical particles in dynamic speckle}
		We simulate the motion of classical, noninteracting point particles in a dynamic, homogeneous speckle potential ${V=V(x,y,t)}$ in two spatial dimensions. To this end, we numerically solve Newton's equation of motion
		\begin{equation}
			m\bm{a} = - \bm{\nabla} V,
		\end{equation}
		where $\bm{a}$ is the acceleration, using the explicit third-order Runge-Kutta method~\cite{numericalmethods}. For the spatial and temporal discretization, we choose ${\Delta x = \Delta y = \SI{100}{\nano\meter}}$ and ${\Delta t = \SI{1}{\micro\second}}$, which are far below all other relevant length and time scales. The simulation encompasses a rectangular region with size ${\SI{22.5}{\micro\meter} \times \SI{22.5}{\micro\meter}}$ that is confined by hard walls. A typical simulation calculates the trajectories of ${\sim \num{50000}}$ particles which start at random positions with velocities drawn from a thermal distribution. Our main observable is the growth rate of the  ensemble-averaged kinetic energy, from which we get the heating rate.
		
		We use a simple numerical approach to simulate a homogeneous two-dimensional speckle pattern. The scalar electric field distribution of a speckle is readily obtained from the discrete fast Fourier transform $\mathcal{F}(R)$ of a two-dimensional square array $R$ filled with random phase factors~\cite{Goodman2007}. Thus, each entry $(k,l)$ of $R$ is given by ${R_{k,l}=\exp\left(2\pi\iu Q\right)}$, where $Q$ is a continuous random variable being uniformly distributed in the interval $[0, 1)$. $R$ represents the electric field of the light after passing through the diffusers. In order to increase the smoothness of the output of $\mathcal{F}$, $R$ is zero-padded. Since we are interested in the speckle intensity distribution $S$, we calculate ${S=\left|\mathcal{F} \left(R\right) \right|^2}$.
		
		Such a static speckle is rendered dynamic by the following procedure. We call $R(t)$ and $S(t)$ the random phase array and corresponding intensity distribution at time $t$. $R(t)$ is propagated in time by adding a small phase $2\pi\iu Q\sqrt{\deltats/\tauc}$ to each entry, where ${\deltats < \tauc}$ is the time step. This simulates the continuous phase evolution on a time scale $\tauc$ that is caused by the rotating diffuser. It is captured by the iteration formula
		\begin{equation}
			R_{k,l}(t+\deltats)=R_{k,l}(t)\times\exp\left(2\pi\iu Q\sqrt{\frac{\deltats}{\tauc}}\right).
			\label{eq:randomarraypropagation}
		\end{equation}
		In order to minimize computational effort, we choose ${\deltats=\tauc / 10 \gg \Delta t}$ and use pointwise linear interpolation between $S(t)$ and $S(t+\deltats)$ for intermediate times. It is important to note that \Cref{eq:randomarraypropagation} does not produce a sequence of speckle patterns $S(t)$ with a correlation time that is precisely given by $\tauc$. The exact value depends on the size of $R$, the zero-padding of $R$, and the choice of $\deltats$ and typically misses $\tauc$ by several \SI{10}{\percent}. Hence, we extract the correlation time from each sequence $S(t)$ by evaluating the auto-correlation function~\cite{gubner2006} of $S(t)$.
			
	\subsection{Derivation of \Cref{eq:heatingrate}}
	To compute the heating rate in the thermal case, we assume the limiting case ${\boltzmann T \gg \vavg}$, which is realized in the experiment. Here, single particles in two dimensions travel on almost straight trajectories through the time-dependent potential. Each time a particle with momentum $p$ traverses a speckle grain, it experiences a "kick", changing its momentum by an amount ${\Delta p \ll p}$ that is proportional to the change in potential height during flyby~\cite{supps}. Due to the random spatial distribution and height of the grains, the particle experiences a series of kicks in random directions, performing a random walk~\cite{pearsonProblem1905}.
				\begin{figure}
			\includegraphics{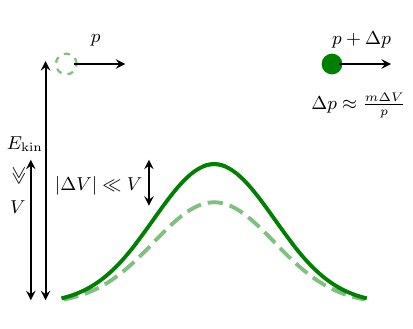}
			\caption{Schematic illustration of a particle traversing a single grain of the dynamic speckle.}
			\label{fig:supp_figure2}
		\end{figure}
		
		
		Quantitatively, consider a particle with mass $m$ and momentum $p$ traveling through the dynamic speckle potential. We make two assumptions concerning the magnitude of $p$.
		\begin{enumerate}
			\item The kinetic energy $\ekin=p^2/(2m)$ of the particle greatly exceeds the average disorder potential $\vavg$. This means that $\boltzmann T \gg \vavg$ for a thermal ensemble.
			\item The velocity $v=p/m$ of the particle is much larger than the largest velocity scale $\vs=\sigmac/\tauc$ of the dynamic speckle.
		\end{enumerate}
		First, we investigate the particle traversing a single speckle grain with potential height $V$ (\Cref{fig:supp_figure2}). Since the grain has the width $\sigmac$, it takes the time ${\Delta t=\sigmac/v}$ to traverse it. Due to the dynamics, the potential height changes by a small amount $\Delta V$. Because of assumption 2, we know that $\left|\Delta V\right| \ll V$. Hence, the particle gains or loses the kinetic energy ${\Delta \ekin = \Delta V}$ and the momentum $\Delta p$. The connection between $\Delta \ekin$ and $\Delta p$ turns out to be
		\begin{equation}
			\Delta \ekin = \frac{(p+\Delta p)^2}{2m} - \frac{p^2}{2m} = \frac{2p\Delta p + \left( \Delta p \right)^2}{2m}.
		\end{equation}
		Since ${\left|\Delta \ekin\right| = \left|\Delta V\right| \ll V \ll \ekin}$ and thus ${\Delta p \ll p}$, we can neglect $\left( \Delta p \right)^2$ and write
		\begin{equation}
			\Delta p \approx \frac{m \Delta V}{p}.
			\label{eq:deltap}
		\end{equation}
		Now, we have the change in particle momentum at a single grain of the speckle potential. Since the speckle is random, we have to calculate the disorder average of the change in momentum or kinetic energy. Therefore, we introduce the disorder average $\avg{\cdot}$ and apply it to the change in kinetic energy to get
		\begin{equation}
			\avg{\Delta \ekin} = \frac{1}{2m}(\avg{2p\Delta p} + \avg{\left( \Delta p \right)^2}).
			\label{eq:tavgdekin}
		\end{equation}
		$\avg{\Delta \ekin}$ is the disorder-averaged change in kinetic energy of a single particle passing by a single speckle grain. Because the two-dimensional disorder is isotropic, the same holds for the direction of $\Delta p$. Hence, the first term in \Cref{eq:tavgdekin} vanishes and we are left with
		\begin{equation}
			\avg{\Delta \ekin} = \frac{1}{2m} \avg{\left( \Delta p \right)^2}.
		\end{equation}
		We plug in $\Delta p$ from \Cref{eq:deltap} to get
		\begin{equation}
			\avg{\Delta \ekin} = \frac{m}{2p^2} \avg{\left( \Delta V \right)^2}.
			\label{eq:tavgdekinsimpl}
		\end{equation}
		Now, we have to evaluate $\avg{\left( \Delta V \right)^2}$. For a given grain with height $V$, the change in height $\Delta V$ during $\Delta t$ is
		\begin{equation}
			\left|\Delta V\right|= \frac{\Delta t}{\tau} V
		\end{equation}
		and \Cref{eq:tavgdekinsimpl} reduces to
		\begin{equation}
			\avg{\Delta \ekin} = \frac{m \Delta t^2}{2p^2 \tau^2} \avg{V^2}.
		\end{equation}
		Due to the exponential potential probability distribution of the speckle, we find $\avg{V^2}=2\vavg^2$. This leads to
		\begin{equation}
			\avg{\Delta \ekin}= \frac{m \Delta t^2}{2p^2 \tau^2} \bar{V}^2 = \frac{\sigma^2}{\tau^2v^4m} \bar{V}^2.
		\end{equation}
		Now, we have the disorder-averaged change in kinetic energy at a single speckle grain. The particle passes grains with a rate $1/(2\Delta t)$, hence
		\begin{equation}
			P(v) = \deriv{T}{t} = \frac{1}{\boltzmann}\deriv{\ekin}{t} \approx \frac{\avg{\Delta \ekin}}{2\Delta t \boltzmann} = \frac{\sigma \bar{V}^2}{2\tau^2v^3m\boltzmann}.
		\end{equation}
		Here, we have made the assumption that two speckle grains are separated by a typical distance $\sigmac$. As to get the temperature dependence of $P$ we integrate $P(v)$
		\begin{equation}
			P(T) = \int_{0}^{\infty} P(v)p(v)\mathrm{d}v.
		\end{equation}		
		over the two-dimensional Maxwell-Boltzmann distribution ${p(v)=2xv\exp\left(-xv^2\right)}$, with ${x=m/(2\boltzmann T)}$.	Unfortunately, the integrand diverges at ${v=0}$ because ${P(v)p(v)\propto v^{-2}}$. Due to assumption 2, we can cut off the integral at $\vs$ without making too big a mistake. We get
		\begin{equation}
			\int_{\vs}^{\infty} \frac{\exp\left(-xv^2\right)}{v^2} \mathrm{d}v = \frac{1}{2}\sqrt{x}\Gamma\left(-\frac{1}{2},x\vs^2\right),
		\end{equation}
		where ${\Gamma(s,q)=\int_{q}^{\infty}t^{s-1}\exp(-t)\mathrm{d}t}$ is the incomplete gamma function. From assumption 2 it follows that ${x\vs^2\ll1}$ and we can approximate ${\Gamma(s,q)\approx-q^s/s}$~\cite{nisthandbookfunctions} to find
		\begin{equation}
			\frac{1}{2}\sqrt{x}\Gamma\left(-\frac{1}{2},x\vs^2\right) \approx \frac{1}{\vs}
		\end{equation}
		Finally, we get
		\begin{equation}
			P(T) = \frac{\sigma\bar{V}^2}{2\boltzmann \tau^2 m} \int_{\vs}^{\infty} \frac{p(v)}{v^3}\mathrm{d}v = \frac{\bar{V}^2}{2 \boltzmann^2 T \tau}.
		\end{equation}

	\subsection{Adaptions between experimental and theoretical data}
		To ensure that the experimental and theoretical data are comparable, we have to make two adaptions.\\
		
		\textbf{Inhomogeneous distribution of average speckle potential}
			Both the numerical simulation and microscopic model assume a homogeneous speckle. In the experiment, a Gaussian envelope with waist ${w\approx\SI{440}{\micro\meter}}$ modulates the local average of the speckle potential. The cloud is located in the center of this envelope. The inhomogeneity is most pronounced along the long ($y$-) axis of the cloud with density distribution $n(y)$. Locally, the heating rate $P$ is proportional to ${\vavg^2(y)=\vavg^2(0)\exp\left(-y^2/w^2\right)}$. Hence, in the experiment, $P$ is reduced by a factor of
			\begin{equation}
				\gamma_1 = \frac{\int n(y) \vavg^2(y) \mathrm{d}y}{\vavg^2(0) \int n(y) \mathrm{d}y}
			\end{equation}
			as compared to the homogeneous case. {\mf Since $\gamma_1$ depends on the precise shape of the density distribution, which is different for the individual data sets, it is computed for each data set independently; it takes on values between \num{.77} and \num{0.93}.}\\

		\textbf{Dimensionality and degrees of freedom}
			As the numerical simulation and microscopic model employ two-dimensional systems and do not include the harmonic trapping potential, the number of degrees of freedom is different from the experiment. In the theory calculations, we have ${d_\mathrm{theo}=2}$ degrees of freedom, assuming we can neglect the weak speckle potential. In the experiment, however, there are ${d_\mathrm{exp}=6}$, two for the harmonic trapping potential and kinetic energy in each dimension. Therefore the additional kinetic energy, as extracted from the numerical simulation and microscopic model, must be equally distributed across $d_\mathrm{exp}$ degrees of freedom. Since the temperature of an ideal gas is ${T=2\ekin/(\boltzmann d)}$, the heating rates of theory and experiment are connected by
			\begin{equation}
				\left(\deriv{T}{t} \right)_\mathrm{exp} = \gamma_2 \left(\deriv{T}{t} \right)_\mathrm{theo},
			\end{equation}
			where ${\gamma_2=\frac{d_\mathrm{theo}}{d_\mathrm{exp}}}=1/3$.\\

		All plotted heating rates from the numerical simulation are corrected by the factor ${\gamma=\gamma_1\gamma_2}$.
		
	\subsection{Direct excitation of superfluid atoms}
	{		In the following we present a theoretical model, which enables us to quantify a direct excitation rate from the superfluid ground state into the thermal cloud due to the speckle potential. We model the gas in local density approximation as a homogeneous ground state with bogoliubov excitations. Using total number conserving bogoliubov theory \cite{Castin2001} enables us to keep the number of condensed atoms $\hat{N}_0 = \hat{a}_0^\dagger \hat{a}_0$ as an operator, and therefore quantify a process which changes the superfluid fraction. To do so we transform the annihilation (creation) operator $\hat{a}_\bk^{(\dagger)}$ of an atom with momentum $\bk$ into operators $\hat{b}_\bk^{(\dagger)}$ describing the annihilation (creation) of a bogoliubov phonon, via
		\begin{align}
		\hat{a}_0^\dagger\; \hat{a}_\bk /\sqrt{N} = u_k \hat{b}_\bk + v_{-k} \hat{b}_{-\bk}^{\dagger},
		\end{align}
	where $u_k$ and $v_k$ are the bogoliubov eigenvectors  \cite{Castin2001} and $N$ is the total number of atoms. The Hamiltonian describing the gas is approximately diagonal in this bases $\hat{H}_0 = \sum_{k\neq0} \hbar \omega_k \hat{b}_\bk^\dagger  \hat{b}_\bk $, where $\omega_k = ck \sqrt{1+k^2\xi^2/2}$ is the bogoliubov dispersion, with the healing length $\xi$ and the speed of sound $c$. The influence of the speckle potential is given by the term
		\begin{align}
		\hat{H}_s &= \sum_{\bk,\bk^\prime} V_{\bk-\bk^\prime}(t)\, \hat{a}_{\bk^\prime}^\dagger \hat{a}_\bk \notag \\
		&= \hat{N}_0 V_0(t) + \sqrt{N} \sum_{\bk \neq 0} V_\bk(t) W_k \; \left( \hat{b}_k + \hat{b}_{-k}^\dagger \right) + \mathcal{O}\left( \hat{b}_k ^2 \right),
		\end{align}
		where $W_k = u_k + v_k$ is a structure factor and $V_\bk(t)  = \int \frac{d^3 r}{L^3} \,  V(\br,t) \;e^{i \bk \br}$ the Fourier transformed speckle potential. We assume that the potential is a Gaussian random variable in space and time, with mean and variance 
		\begin{align}
		\overline{V(\br,t)} &= 0\\ \quad \overline{V(\br,t)V(0,0)} &=  \eta^2 \bar{V}^2 \exp \left(-\frac{\br^2}{\sigma^2}-\frac{t^2}{\ttau^2}  \right),
		\end{align}
		where $\eta$ is a fitting factor which we need to describe the trapped system with a theory of a homogeneous gas and $\ttau = \tau/\sqrt{\log{2}}$. 
		The Heisenberg equation of motion of $\hat{b}_\bk$ and $\hat{N}_0$ are given by
		\begin{align}
		\frac{d}{dt} \hat{b}_\bk = -i\left(\omega_k \hat{b}_\bk + \frac{1}{\hbar} V_{-\bk}(t) W_k \sqrt{N} \right)  \\
		\frac{d}{dt} \hat{N_0} = \frac{i}{\hbar} \sqrt{N} \sum_{\bk \neq 0}   V_\bk(t) W_k^{-1} \left(\hat{b}_{-\bk}^\dagger - \hat{b}_\bk\right),
		\end{align}
		where terms which do not scale with $\sqrt{N}$ where neglected. These equations can be integrated out exactly and we find for the averaged expectation value of condensed particle number $N_0 = \overline{\langle \hat{N_0} \rangle}$
		\begin{align}
		&N_0(t) -N_0(0) && \notag \\
		=& - 2 \frac{N}{\hbar^2}   \; \text{Re}\bigg( \sum_{k\neq 0} \, \int_0^t dt^\prime \; \int_0^{t^\prime} dt^{\prime \prime} \; e^{i \omega_k (t^\prime-t^{\prime\prime})} \notag \\
		& \quad \quad \cdot \overline{ V_{\bk}(t^\prime) V_{-\bk}(t^{\prime \prime}) } \bigg).
		\end{align}
		This simplifies for late times $t \gg \tau$ to a linear excitation rate of ground state atoms 
		\begin{align}
		N_0(t) - N_0(0) =  - N \; \Gamma \; t,
		\end{align}
		where the transition rate is given by
		\begin{align}
		\Gamma =& \frac{ \eta^2 \bar{V}^2\;  \pi \sigma^3 }{3 \hbar^2 v_c  \xi\;\sqrt{\sigma^2+v_c^2 \ttau^2  }}  e^u \bigg\{ 2u \Big[ I_{5/4}(u) \notag \\
			&  -I_{3/4}(u)+I_{1/4}(u) -I_{-1/4}(u) \Big] +I_{1/4}(u) \bigg\}, 
		\end{align}
		Here $I_{\nu}(u)$ are modified Bessel function of the first kind \cite{abramowitz2008} and $u$ is defined as
		\begin{align}
		u = \frac{(\sigma^2 + v_c^2 \ttau^2)^2}{16 \xi^2 v_c^2 \ttau^2}.
		\end{align}
		At this point we assume, that the gas thermalized quickly via internal scattering, such that the decrease of condensed atoms directly leads to an increase in temperature. This results in an additional heating rate, which we calculate from the leading order non interacting part of the superfluid fraction $N_0/N = 1-(T/T_c)^3$ and find
		\begin{align}
		\tNo = \frac{T_c^3}{3T^2} \Gamma.
		\end{align}		
}
	\subsection{Rate model for the dissipation of BECs}
\begin{figure}[b]
			\includegraphics{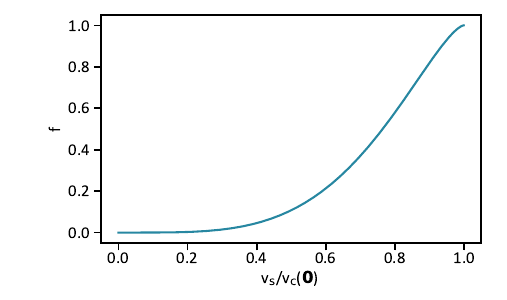}
			\caption{Fraction $f$ of particles located in regions of the condensate where ${\vc(\bm{r})<\vs}$ versus $\vs/\vc(\bm{0})$.}
			\label{fig:supp_figure3}
		\end{figure}

		In the following we give a detailed description of the rate model
\begin{align}
			\dot{N} &= \nexdot + \nmoldot \label{eq:ratemodelsupps1} \\
			\dot{T} &= \ts + \tev \label{eq:ratemodelsupps2},
\end{align}
		for the total particle number ${N=\nc+\nex}$ and the temperature $T$. The number of particles in the superfluid ${\nc=N\times \cf}$ is given by the superfluid fraction $\cf$. $\cf$ coincides with the condensate fraction, provided that quantum depletion is negligible. Otherwise, $\cf$ exceeds the condensate fraction. In order to account for the interaction between particles and the finite size of the system, we solve the transcendental equation			
		\begin{widetext}
			\begin{align}
				\cf &= \overbrace{1 - \left(\frac{T}{\tc}\right)^3}^\text{noninteracting} - \overbrace{\frac{\zeta(2)}{\zeta(3)} \left(\frac{T}{\tc}\right)^2 \left( \left(1+0.16 \eta^3\cf^{1/5}\right) \eta\cf^{2/5} \right)}^\text{interaction corrections} - \overbrace{\frac{3\omega_\mathrm{a}\zeta(2)}{2\omega_\mathrm{g}\zeta(3)^{2/3}} \left(\frac{T}{\tc}\right)^2 N^{-1/3}}^\text{finite-size correction}
				\label{eq:xi}
			\end{align}
		\end{widetext}
		to determine $\cf$~\cite{xiongCritical2001}. The first term of \Cref{eq:xi} is the well-known result for a noninteracting gas in a harmonic trap that only depends on $T$ and the critical temperature 
		\begin{equation}
			\tc = \frac{\hbar\omega_\mathrm{g}}{\boltzmann}\left(\frac{N}{\zeta(3)}\right)^{1/3},
		\end{equation}
		where $\hbar$ is the reduced Planck constant, $\omega_\mathrm{g}$ the geometric mean of the trapping frequencies, and $\zeta$ the Riemann zeta function. The second term includes a first-order correction due to interactions, quantified by the dimensionless parameter	
		\begin{equation}
			\eta = \frac{1}{2}\zeta(3)^{1/3}\left(15 N^{1/6} \frac{a}{a_\mathrm{ho}}\right)^{2/5}
		\end{equation}
		with the oscillator length ${a_\mathrm{ho}=\sqrt{\hbar/(m\omega_\mathrm{g})}}$, and the Lee-Huang-Yang correction~\cite{leeEigenvalues1957}. The third and last term is the finite-size correction with $\omega_\mathrm{a}$ the arithmetic mean of the trapping frequencies. Disorder-induced depletion of the condensate fraction is negligible in our system, because the healing length $\xi$ is roughly a factor two below the smallest correlation length~\cite{abdullaevBoseEinstein2012}. After solving \Cref{eq:xi}, $\cf$ is reduced by the fraction $f$ of particles that are located in a region of the condensate density $n_0(\bm{r})$ where the local critical velocity $\vc(\bm{r})$ is below the largest velocity scale $\vs$ of the dynamic speckle (see \Cref{fig:supp_figure3}). Hence, ${\cf \rightarrow \cf \times (1-f)}$ with		
		\begin{equation}
			f = \frac{1}{\nc} \int_{\vs>\vc(\bm{r})} n_0(\bm{r}) \dder r^3,
		\end{equation}
		where $n_0(\bm{r})$ is the Thomas-Fermi density distribution.
		The heating of the thermal fraction is described by ${\ts=P(t)\gamma}$. We get the heating rate $P(t)$ from the numerical simulation and incorporate the time dependence of $\vavg$ as shown in \Cref{fig:supp_figure1}. Evaporation from the thermal fraction is captured by
		\begin{align}
			\nexdot &= -\frac{\nex}{\tevap} \\
			\tev   &= -\frac{1}{\tevap} \left(\frac{U_0}{3 \boltzmann} - T\right),
		\end{align}
		where $1/\tevap$ is the evaporation rate and ${U_0=\SI{438}{\nano\kelvin}\times\boltzmann}$ the trap depth. The evaporation rate
		\begin{equation}
			\frac{1}{\tevap} = \frac{1}{\tcoll} \frac{U_0}{\sqrt{2} \boltzmann T} \exp\left(-\frac{U_0}{\boltzmann T}\right)
		\end{equation}
		depends on the elastic scattering rate $1/\tcoll$ and the probability of collision events which leave one of the particles in a state with energy $>U_0$~\cite{pethick_smith_2008}. We calculate the scattering rate
		\begin{equation}
			\frac{1}{\tcoll} = \bar{v}_\mathrm{rel} \sigma_\mathrm{coll}\left(n_0^\mathrm{max}+ n_\mathrm{th}^\mathrm{max}\right)
		\end{equation}
		from the average relative velocity ${\bar{v}_\mathrm{rel} = \sqrt{2}\sqrt{8\boltzmann T/(m\pi)}}$ of a thermal gas in three dimensions, the scattering cross section ${\sigma_\mathrm{coll} = 8\pi a^2/(1+k_\mathrm{dB}^2 a^2)}$ for indistinguishable particles with the thermal de Broglie wave vector ${k_\mathrm{dB}=\sqrt{2\pi m \boltzmann T}/\hbar}$, and the peak densities of the BEC ${n_0^\mathrm{max} = \mu/g}$ and the thermal cloud ${n_\mathrm{th}^\mathrm{max} = \nex \left(m \omega_\mathrm{g}^2/(2\boltzmann T\pi)\right)^{3/2}}$~\cite{pethick_smith_2008}. At last, we include molecular relaxation $\nmoldot$, which is a two-body process and hence described by the differential equation
		\begin{equation}
			\dot{n} = -\alphamol n^2,
			\label{eq:molrelax1}
		\end{equation}
		with the rate of molecular relaxation $\alphamol$. For simplicity, we treat the density of the thermal and condensed clouds separately by writing
		\begin{equation}
			\dot{n} \approx \dot{n}_0 + \dot{n}_\mathrm{th} = - \alpha (n_0^2 + n_\mathrm{th}^2)
			\label{eq:molrelax2}
		\end{equation}
		Integration of \Cref{eq:molrelax2} over all space yields
		\begin{equation}
			\nmoldot = -\alpha \left(\frac{4}{7} n_0^\mathrm{max} \nc + \frac{1}{2\sqrt{2}} n_\mathrm{th}^\mathrm{max}\nex  \right),
		\end{equation}
		where we have assumed a Gaussian density distribution of the thermal molecules~\cite{pethick_smith_2008}. The determined loss rates $\alphamol$ are given in \Cref{tab:parameters}.
		
		Employing Wolfram Mathematica, we solve \Cref{eq:ratemodelsupps1,eq:ratemodelsupps2} numerically with initial conditions ${T(0)=\SI{35}{\nano\kelvin}}$ for all measurement series and $N(0)$ as extracted from absorption images at ${\ds=0}$ and ${\tauc=\infty}$ for each respective measurement series~(see \Cref{tab:parameters}). The initial temperature is adjusted such that evaporation is negligible during the experimental sequence~(\Cref{fig:supp_figure1}) with no speckle potential present and is well within the margin of error of the experimentally determined temperature of $\SI{50(25)}{\nano\kelvin}$.
		
		The failure of our phenomenological model to explain the experimental data for reduced interaction strength can be explained by the following argument. Exceeding the superfluid critical velocity, elementary excitations are created with energy ${E<\mu}$, which cannot remove particles from the trap. The decay of such excitations is possible only via interaction with other thermal or disorder-induced excitations, leading to the formation of a higher-energy excitation, which can eventually remove molecules from the trap. The corresponding damping rate of such excitations has been shown to scale with the interaction parameter as $(n_0 a^3)^{1/2}$~\cite{fedichevDamping1998}. Thus, for large scattering length, the damping is sufficiently fast to justify the assumption of an immediate depletion of the superfluid density. For weaker interaction, by contrast, the damping rate does not suffice to cause immediate particle loss, and our model overestimates the loss rate. Furthermore, as the local density approaches zero in the outer regions of the condensate, the healing length grows, and the local chemical potential is diminished, effectively shielding the BEC against the disorder evolution on short time and length scales~\cite{sanchez-palenciaSmoothing2006a}.

\end{document}